\documentclass{phb-proc4-auth}

\usepackage{graphicx}
\usepackage{amssymb}
\usepackage{times}

\begin{document}
\begin{frontmatter}


\journal{SCES '04}


\title{Characterisation of Anderson localisation using distributions}

%
%
%
%
%
%

\author[HGW,RRZE]{A. Alvermann\corauthref{1}}
\author[HGW]{G. Schubert}
\author[SYD]{A. Wei{\ss}e}
\author[HGW]{F. X. Bronold}
\author[HGW]{H. Fehske}

%
 
\address[HGW]{Institut f\"ur Physik, Ernst-Moritz-Arndt-Universit\"at
  Greifswald, Germany}
\address[RRZE]{Regionales Rechenzentrum Erlangen, Universit\"at
  Erlangen, Germany}
\address[SYD]{School of Physics, The University of New South Wales, 
Sydney, Australia}

%
%
%
%


%
%
%
%

\corauth[1]{Corresponding Author: Institut f\"ur Physik,
  Ernst-Moritz-Arndt-Universit\"at Greifswald, 17489 Greifswald,
 Phone: +49-(0)3834-864765, Fax: +49-(0)3834-864701,
 Email: alvermann@physik.uni-greifswald.de}


\begin{abstract}
We examine the use of distributions in numerical treatments of
Anderson localisation
and supply evidence
that treating exponential localisation on Bethe lattices 
recovers the overall picture known from
hypercubic lattices in 3d.
\end{abstract}

%
%

\begin{keyword}
disordered electron systems \sep metal-insulator transition
\end{keyword}


\end{frontmatter}

%
%
%
%
%

The first question when studying localisation of particles
is how to characterise localised states in distinction to extended
states.
 Concepts as the localisation length or
return probability are  intuitive quantities, but
restricted to non-interacting particles,
and quantities as the conductance are hard to calculate, at
least for disordered interacting systems.

Recalling Anderson's first ideas about localisation~\cite{and}
a characterisation using distributions of
the local density of states (LDOS) $\rho_i$ as the basic quantity
is appealing
since 
such a description can be adopted to interacting systems as well.
This 
leads us to consider two different approaches to the
localisation problem given by the Anderson model
$ H= \sum_i \epsilon_i c^\dagger_i c_i
 + t \sum_{\langle i,j \rangle} c^\dagger_i c_j$
 with box distribution
$P(\epsilon_i)=\Theta(\gamma/2-|\epsilon_i|)/\gamma$.

\begin{figure}
\centering
\includegraphics[width=0.4\textwidth]{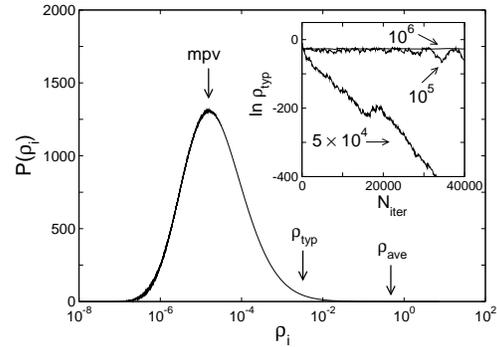}
\caption{
Distribution of LDOS for strong disorder 
(band center $\omega=0.0$, 
$\gamma=2.0 \approx 0.66 \times \gamma_\mathrm{crit}$).
Arrows indicate the most probable value (mpv), the typical density of
states $\rho_\mathrm{typ}$, and the average density of states
$\rho_\mathrm{ave}$.
The inset displays 
$\rho_\mathrm{typ}$ nearly at the localisation transition
($\omega=0.0$, $\gamma=3.0$)
 versus the number of iterations $N_{iter}$ 
in the Monte-Carlo procedure,
  for samples sizes $5\times 10^4, 10^5, 10^6$ from bottom to top.
Increasing the sample size allows to separate localised and extended
states arbitrarily close to the localisation transition.
}
\end{figure}

The AAT-scheme~\cite{abou} 
formulates the localisation problem directly in terms of
distributions.
It provides a selfconsistency equation for the
distribution of the local (retarted) Greensfunction $G_{ii}$
which can be solved via Monte-Carlo-sampling.
By its construction on a Bethe-lattice, being a loop-free
approximation to hypercubic lattices,
AAT can be understood as a kind of mean field approach to the
localisation problem.
Clearly, neglecting all nontrivial loops prevents any treatment of
weak localisation.
In contrast, exponential localisation can be treated to a large
extent, as shall be demonstrated here.
To compare with
localisation on hypercubic lattices a refinement of the kernel
polynomial method (KPM) serves as a quasi-exact
numerical procedure to construct the LDOS-distribution without
calculating wavefunctions of the disordered system,
hence reducing both computational time and memory demands
(for details see~\cite{schub}).

In Fig. 1 the distribution of the LDOS  for strong disorder is
shown.
The appearance of heavy tails 
and a high peak at small values indicates 
the qualitative behaviour of the distribution at the onset of
localisation.
Increasing the disorder, thus pushing the system over the localisation
 transition,
 causes the distribution to be singular, being entirely concentrated
 at $\rho_i=0$.
Evidently the arithmetic mean $\rho_\mathrm{ave}=\langle\rho_i\rangle$
 is inappropriate to detect this qualitative change.
But other quantities, e.g. the most probable value or the geometric mean
(``typical density of states''
$\rho_\mathrm{typ}=\exp\langle\ln\rho_i\rangle$),
 that put sufficient weight on
small values of $\rho_i$, will be critical at the transition. 
We have discussed elsewhere~\cite{baf} how to use the typical density
of states in combination with a 
thorough investigation of its limiting behaviour close to the real
energy axis
to extract 
the position of the localisation transition from the distribution
(which allowed for a calculation of mobility edges in an
\emph{interacting} system).

\begin{figure}
\centering
\includegraphics[width=0.4\textwidth]{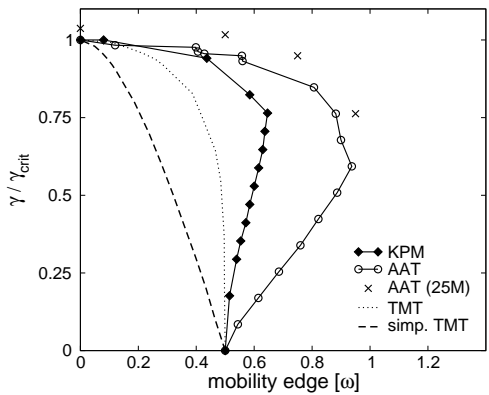}
\caption{
Mobility edge trajectory for the Anderson model
calculated with KPM for a 3d-lattice (diamonds), 
and with AAT (circles),
representing the distribution through a sample with $5 \times 10^4$
elements.
The crosses indicate points in the $(\omega,\gamma)$-plane, which
correspond to delocalised states if sampled with $2.5\times 10^7$ elements.
The dotted (dashed) line shows the mobility edge trajectory resulting
from TMT (simplified TMT-variant).
}
\end{figure}

Fig. 2 shows mobility edge trajectories calculated with AAT and KPM.
Both trajectories show the same qualitative features.
In contrast to KPM the AAT scheme does not suffer from limitations due
to the finite geometric size of the system under consideration.
Though the resolution which the distribution is sampled
with in AAT is restricted by the finite size of the Monte-Carlo
sample, this limitation can be easily overcome
by increasing the sample size (cf. Fig. 1, 2).
In principle this allows for an extremely precise determination of
mobility edges within AAT.

Although both KPM and AAT
construct the full distribution of the LDOS
the localisation transition already manifests itself in a suitable
moment thereof
(e.g. the typical density of states).
As a matter of course,
this triggers the question
whether the localisation problem can be treated by an
effective theory using this moment as an (scalar and local) order parameter.

Indeed, the typical medium theory (TMT)~\cite{tmt} which is
entirely based on the typical density of states
is believed to capture the most basic effects of Anderson
localisation, and has been recently applied to the Anderson-Hubbard
model~\cite{voll}.
From a pragmatic point of view it is quite favorable to replace 
AAT by the simpler
TMT.
However, the typical density of states is no obvious choice for the
order parameter.
In contrast to what is sometimes claimed 
it provides no good approximation to the most probable value
(cf. Fig. 1). 
So we tried to simplify even further and replaced 
the typical density of states with the minimal values of the LDOS calculated 
for the maximal value of the on-site potential $\epsilon_i$, i.e. 
$\epsilon_i=\pm \gamma/2$ for the box distribution.
This means to incorporate only the maximal
scattering contribution, originating from the deepest impurity.
We can then calculate a mobility edge trajectory which, at a first
glance, does equally well compared with the one resulting from TMT
(cf. Fig. 2).
Nevertheless, the two approaches fail to obtain the reentrant behaviour
of the mobility edges.
It seems that both TMT and our simplified TMT-variant
do only capture localisation caused by deep impurities
(``pushing states outside the band''),
but not the subtle effects caused by quantum interference.
Consistent with this interpretation the critical disorder predicted by
these approaches is smaller than given by AAT.
Therefore we come to the conclusion 
that the treatment of disordered systems 
within TMT might be inherently problematic.

To summarise:
AAT is a theory of high precision for exponential localisation on the Bethe
lattice
which recovers all qualitative effects of
 exponential localisation on hypercubic lattices.
The use of distributions in numerical treatments
of Anderson localisation is further justified
by a comparison to results obtained for 
3d-hypercubic lattices.
While localisation effects originating from local correlations
are fully contained in AAT 
they are partially lost in averaged descriptions as TMT.
%

%
%
%
%

%
%
%
%



\begin{thebibliography}{00}

\bibitem{and}
P.~W.~Anderson, \emph{Phys. Rev.} \textbf{109}, 1492 (1958).

\bibitem{abou}
R.~Abou-Chacra, P.~W. Anderson, and D.~J. Thouless, \emph{J. Phys. C}
  \textbf{6}, 1734 (1973).

\bibitem{schub}
G.~Schubert, A.~Wei{\ss}e, and H.~Fehske, \emph{cond-mat/0309015}  (2003).


\bibitem{baf}
F.~X.~Bronold, A.~Alvermann, and H.~Fehske, \emph{Phil. Mag.} 
\textbf{84}, 673  (2004).



\bibitem{tmt}
V.~Dobrosavljevi\'{c}, A.~A.~Pastor, and B.~K.~Nikoli\'{c}, \emph{Europhys.
  Lett.} \textbf{62}, 76 (2003).

\bibitem{voll}
K.~Byczuk, W.~Hofstetter, and D.~Vollhardt, \emph{cond-mat/0403765} (2004).

\end{thebibliography}
\end{document}